\documentclass[%
 reprint,
 amsmath,amssymb,
 aps,
]{revtex4-1}

\usepackage{graphicx}% Include figure files
\usepackage{dcolumn}% Align table columns on decimal point
\usepackage{bm}% bold math
\begin{document}

%\preprint{APS/123-QED}

\title{
Sub 100-fs, 5.2-$\mu$m ZGP Parametric Amplifier Driven
by a ps Ho:YAG Chirped Pulse Amplifier and its application to high harmonic generation
}


\author{Tsuneto Kanai}
 \email{goldwell74@gmail.com}
\affiliation{Photonics Institute, Vienna University of Technology, Gusshausstrasse 27-387, A-1040 Vienna, Austria}

\author{Pavel Malevich}
\affiliation{Photonics Institute, Vienna University of Technology, Gusshausstrasse 27-387, A-1040 Vienna, Austria}

\author{Sarayoo Sasidharan Kangaparambil}
\affiliation{Photonics Institute, Vienna University of Technology, Gusshausstrasse 27-387, A-1040 Vienna, Austria}

\author{Kakuta Ishida}
\affiliation{Department of Chemistry, School of Science, The University of Tokyo, 7-3-1 Hongo, Bunkyo-ku, Tokyo 113-0033, Japan}

\author{Makoto Mizui}
\affiliation{Department of Chemistry, School of Science, The University of Tokyo, 7-3-1 Hongo, Bunkyo-ku, Tokyo 113-0033, Japan}

\author{Kaoru Yamanouchi}
\affiliation{Department of Chemistry, School of Science, The University of Tokyo, 7-3-1 Hongo, Bunkyo-ku, Tokyo 113-0033, Japan}

\author{Heinar Hoogland}
\affiliation{Menlo Systems GmbH, Am Klopferspitz 19a, 82152 Martinsried, Germany}
\affiliation{Department of Physics, University of Erlangen-Nuremberg, Staudtstr. 1, 91058 Erlangen, Germany}

\author{Ronald Holzwarth}
\affiliation{Menlo Systems GmbH, Am Klopferspitz 19a, 82152 Martinsried, Germany}

\author{Audrius Pug\v{z}lys}
\affiliation{Photonics Institute, Vienna University of Technology, Gusshausstrasse 27-387, A-1040 Vienna, Austria}
\affiliation{Center for Physical Sciences \& Technology, Savanoriu Ave. 231 LT-02300 Vilnius, Lithuania}

\author{Andrius Baltu\v{s}ka}
\affiliation{Photonics Institute, Vienna University of Technology, Gusshausstrasse 27-387, A-1040 Vienna, Austria}
\affiliation{Center for Physical Sciences \& Technology, Savanoriu Ave. 231 LT-02300 Vilnius, Lithuania}


\begin{abstract}
 We report a 1 kHz repetition-rate mid-IR (MIR) optical parametric amplifier (OPA) system operating
 at a central wavelength of 5.2 $\mu$m with the tail-to-tail spectrum extending over 1.5 $\mu$m
 and delivering 40 $\mu$J pulses that are compressed to
 99 fs (5.6 optical cycles).  Also we develop a novel pulse compression scheme for further pulse compression and wavelength tunability.
 As the first application of this laser system, we generated high harmonics in bulk ZnSe above the bandgap, dense exciton generation after 10-photon absorption, high order sum- and difference-frequency generation,
 ultrafast  transition in the conduction band, which reflects the structure of conduction bands.
%\begin{description}
%\end{description}
\end{abstract}

\maketitle

%\tableofcontents

  High energy femtosecond lasers in the 5-10 $\mu$m region \cite{malevich2016broadband,sanchez20167} attract much attention as next generation lasers for unexplored strong field physics, where longer wavelengths compared to that of typical 0.8 $\mu$m drivers have many advantages as  demonstrated with 3.9-$\mu$m optical parametric chirped pulse amplifier \cite{Andriukaitis2011} driving high harmonic generation (HHG) \cite{popmintchev2012bright}, plasma hard X-ray pulse generation \cite{weisshaupt2014high}, and MIR filamentation \cite{mitrofanov2015mid}.
In terms of the effective order of the wavelength conversion processes, 
laser development in this wavelength region, can be categorized into two types. 
One is based on effectively third order processes such as four wave mixing  \cite{nomura2013frequency},
ZnGeP$_2$ (ZGP)/CdSiP$_2$ (CSP) OPA driven by a ps Yb:YAG CPA-pumped 2$\mu$m OPCPA  \cite{liang2016octave},
ZGP OPA driven by a Ti:Sa CPA-pumped 2$\mu$m OPA pumped  \cite{wandel2016parametric},
AgGaS$_2$ (AGS)-based difference frequency generation (DFG)
between signal and idler from  Ti:Sa CPA-pumped BBO OPA  \cite{golubovic1998all,Lanin2014}.  
While typical conversion efficiency of this type is as low as $\approx 1$  \% as is understood by the high order of their nonlinearity,
one can directly obtain ultrashort 5-10 $\mu$m pulses, whose pulse duration is comparable to those of pump lasers.
The other type is based on purely second order OPAs pumped by CPA lasers such as a Ho:YAG CPA-driven ZGP OPA \cite{malevich2016broadband},
a Ho:YLF CPA-driven ZGP OPA \cite{sanchez20167}, Ti:Sa laser-pumped LiGaS$_2$ OPA \cite{petrov2004second}, LiInS$_2$ OPA \cite{rotermund2001optical},
CdHgGaS OPA  \cite{petrov2004mid}, and GaSe DFG between the broad gain spectrum of Ti:Sa \cite{kaindl1999broadband}.
The advantage of this method is in its reasonable high conversion efficiency $\approx 20$ \%
and therefore these designs have a capability for scaling up their pulse energy/power by choosing suitable crystals such as ZGP and CSP.
On the other hand, one needs special picosecond pump lasers whose wavelength are long (typically $>2$ $\mu$m) enough to avoid one- or two-photon absorption,
which is usually present in the user friendly wavelength region of 0.8-1 $\mu$m  
due to the relatively small bandgap of these MIR nonlinear crystals.
Also the behavior of such MIR nonlinear semiconductor crystals for high energy femtosecond or picosecond pulses is in unexplored parameter space.

In our previous publication \cite{malevich2016broadband}, which is categorized into the latter type, 
we report on generation of 80 $\mu$J pulses at 5.3 $\mu$m  
via ZGP OPA pumped by a Ho:YAG picosecond CPA, while its compression and measuring its pulse duration were
left as next tasks.  Main difficulty was partly observed large and complicated pulse distortion, which is not present for typical visible/near-IR femtosecond OPA nor nanosecond MIR OPA.
Also the previous design is not capable of stabilizing carrier envelope phase (CEP) of the MIR pulses.
In this paper, we clarify its underlying physics as cascaded $\chi^2$ process \cite{desalvo1992self} and huge third order dispersion (TOD) by humidity in the air 
and based on this revealed physics, we demonstrate two approaches for the generation of femtosecond pulses centered at 5.2 $\mu$m,
which are characterized by a newly developed MIR second harmonics generation frequency resolved optical gating (SHG FROG) setup.
One of the methods relies on the amplification of broadband seed pulses while avoiding a back-conversion of the signal and idler into the pump.
Another method is based on the amplification of relatively narrow-band seed with simultaneous broadening of the spectrum due to the phase modulation induced by the cascaded $\chi^2$ process \cite{desalvo1992self}.
SHG FROG measurements reveal that generated pulses are compressible by controlling second and third order phase, which makes the second method advantageous since it allows ones to tune the center wavelength to the spectral region where broadband seed cannot be generated.
Furthermore the system is modified for stabilizing CEP of both signal and idler pulses in the MIR region and stabilizing CEP of seed is demonstrated
by a novel filamentation-based 2$f$-3$f$ interferometer.
Finally, as an application of the generated femtosecond MIR pulses, we demonstrate generation of high harmonics in ZnSe with the spectrum extending above its bandgap.
Observed high harmonic spectra serve as a novel tool for ultrafast probing of band-structure in dielectrics and semiconductor materials \cite{luu2015extreme,ghimire2011observation,vampa2015linking,Chin2001}
as well as for stabilizing CEP of MIR pulses to develop \textit{perfect wave} setup \cite{haessler2015enhanced} in the MIR region.
%
%%%%%%%%%%%%%%%%%%%%%%%%%%%%%%%%%%%%%%%%%%
\begin{figure}[htbp]
 \centering
 \includegraphics[width=\linewidth]{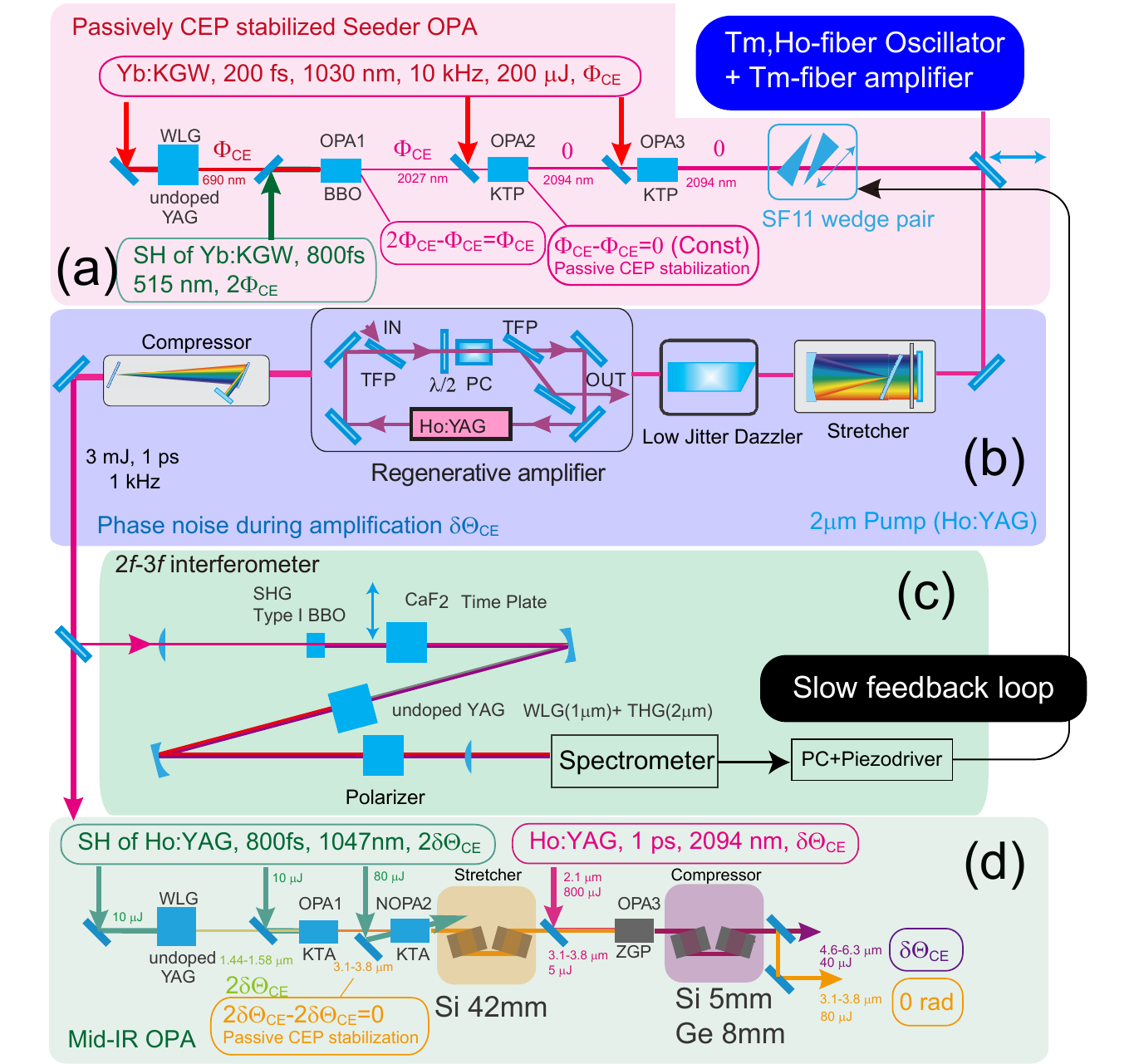}
 \caption{%
 \footnotesize{%
 Optical layout consisting of (a) a passively CEP stabilized seeder OPA, (b) a CEP stabilized 2 $\mu$m pump (Ho:YAG),
 (c) a $2f$-$3f$ interferometer, and (d) a CEP stabilized MIR OPA.
 }
 }
\end{figure}
%%%%%%%%%%%%%%%%%%%%%%%%%%%%%%%%%%%%%%%%%%
%
%
%%%%%%%%%%%%%%%%%%%%%%%%%%%%%%%%%%%%%%%%%%%%%
%\section{laser}
%%%%%%%%%%%%%%%%%%%%%%%%%%%%%%%%%%%%%%%%%%%%%
%
%

Figure 1  shows our optical layout consisting of (a) a seeder OPA \cite{hoogland2014fiber}, (b) a CEP stabilized 2.09 $\mu$m pump (Ho:YAG),
(c) a $2f$-$3f$ interferometer, and (d) a CEP stabilized MIR OPA.
Basic design except (c) is described in Ref.~\cite{hoogland2014fiber} and here we restrict ourselves to its brief description and focus on the main modified points for the present purposes.
For the generation of 2.09-$\mu$m seed for Ho:YAG CPA system we developed two sources, namely a Tm,Ho-fiber oscillator/power amplifier seeder \cite{hoogland2014fiber}, which assures stable and hands-free operation, and an Yb:KGW CPA (Pharos, Light conversion Ltd.)-driven KTiOPO$_4$ (KTP) OPA \cite{malevich2013high}, which has advantages related to higher-energy and passive CEP-stabilization \cite{baltuvska2002controlling} of seed pulses.
As is shown in Fig.~1(a) and Fig.~2, passive CEP-stabilization is achieved through a DFG process at the second OPA stage.
Before amplification, 2.09-$\mu$m seed is stretched to 140 ps pulse duration with a grating based Martinez stretcher containing a spectrum shaper which helps to compensate a  gain narrowing
during amplification.
In order to handle higher-order dispersion we employ an acousto-optic programmable dispersive filter [AOPDF, (Dazzler, Fastlite)] operated in a low jitter mode for reducing the phase noise.
Stretched seed is amplified in a ring-cavity regenerative amplifier (RA) based on a water cooled Brewster cut Ho:YAG crystal, which is pumped by a 100 W, 1907-nm Tm-fiber laser (TLR-100-WC-Y12, IPG Photonics).
The RA is capable of generating 9.5-mJ pulses before compression at 1 kHz repetition rate.
In the present experiments, however, in order to reduce the B-integral and to maintain good spatial quality and compressibility of generated pulses, the amplifier was operated in the regime of moderate amplification.
At a repetition rate of 1 kHz, 5 mJ pulses were extracted from the RA at pump power of 30 W and compressed to 1 ps pulse duration in a Treacy compressor having transmission efficiency of 60\%.

Generated 2.09–$\mu$m pulse were employed to drive a ZGP OPA (Fig.~1(d)) operating at 5.2 $\mu$m central wavelength.
Since it is rather difficult to produce seed in the vicinity of either 5.2 $\mu$m (idler wave) or 3.5 $\mu$m (signal wave) directly from 2.09 $\mu$m, we opted for cascaded seed generation.
For the generation of white light (WL), a portion of 2.09 $\mu$m light was frequency doubled in a BBO  crystal.
Because of the nonlinear conversion, second harmonic pulses are shortened to 800 fs, which allows stable WL generation in the vicinity of 1.5 $\mu$m in a 10-mm long YAG crystal.
Amplification of the WL in a 10-mm-long type II KTA crystal ($\theta=41^\circ, \phi=0^\circ$), pumped by the remaining 1047-nm light,
results in the generation of CEP-stable idler pulses in the vicinity of 3.5 $\mu$m.
The 3.5-$\mu$m pulses were amplified in the second stage OPA (10-mm-long Type II KTA, $\theta=39^\circ$, $\phi=0^\circ$, 
noncollinear angle $\Delta \theta = 2^\circ$) to typically 5 $\mu$J.
In the third stage, containing a type I ZGP ($\theta=53^\circ$, 4-mm or 8-mm long) pumped at 2.09 $\mu$m, the 3.5-$\mu$m pulses were injected as seeds, 
resulting in the generation of 50 $\mu$J, 5.2-$\mu$m idler pulses.  Here as shown in Fig.~1(d), CEP of the 5.2 $\mu$m pulses becomes the same as that of 2.09 $\mu$m pulse and the stabilizing it is crucial.
%
%%%%%%%%%%%%%%%%%%%%%%%%%%%%%%
\begin{figure}[htbp]
 \centering\includegraphics[width=\linewidth]{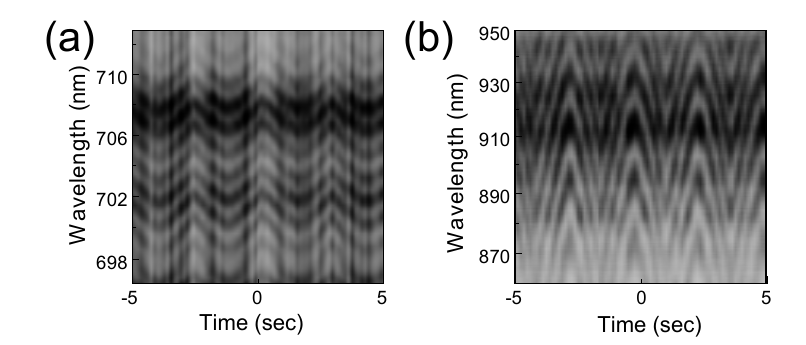}
 \caption{%
 \footnotesize{%
 Controlled CEP fringe measured with a $2f$-$3f$ interferometer.  (a) Unstable typical CEP fringes around 700 nm with time plate (12-mm-thick CaF$_2$ plate)
 and (b) stable CEP fringes around 900 nm without time plate controlled by a SF11 wedge pair on a PC controlled piezo stage.
 This stability is from a stable two-color filamentation by 2.09-$\mu$m pulses and their SH pulses, which have pulse energy of 2.1 $\mu$J and 1.2 $\mu$J, respectively.
 }
 }
\end{figure}
%%%%%%%%%%%%%%%%%%%%%%%%%%%%%

For measuring and stabilizing CEP of the picosecond 2 $\mu$m laser, we adopted a $2f$-$3f$ interferometer,
where WL was obtained after the pulse shortening during SH process of 2.09 $\mu$m pulses again and interfere with the third harmonic (TH) of 2.09 $\mu$m pulses.
Figure 2 (a) shows a measured $2f$-$3f$ interferogram where a saw-tooth modulation of CEP was applied by a SF11 wedge pair (Fig.~1(a)).
Wavelength was around 698 nm and to separate the TH of 2.09 $\mu$m pulses and the WL temporally, time plate (12-mm-thick CaF$_2$ plate) was inserted.
The observed instability is from that of THG in the single color filament by the 2.09 $\mu$m pulses and was removed by taking off the time plate so to enter two color filamentation regime,
which was stabilized by the short SH of 2.09 $\mu$m pulses (Fig.~2(b)).  Also for amplified ps pulses,
this stabilization effect was observed while their CEP noise was larger than the stabilized CEP of seed pulses ( $\approx 200$ mrad).
Interestingly in this regime, we found the most stable wavelength region was around 900 nm rather than the that of the TH (698 nm),
which will be discussed in the near future as well as CEP of MIR pulses although CEP fringes were observed in Fig.~5(b).
%
%%%%%%%%%%%%%%%%%%%%%%%%%%%%%%%%%%%%%%%%%%
\begin{figure}[htbp]
 \centering
 \includegraphics[width=\linewidth]{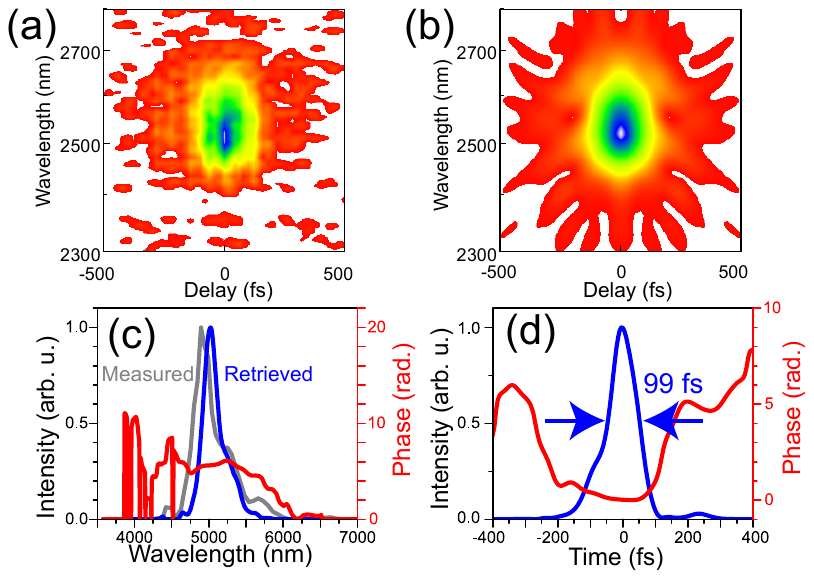}
 \caption{%
 \footnotesize{%
 SHG-FROG characterization of the 99-fs, 5.2-$\mu$m pulses generated in a 4-mm-thick ZGP crystal.  Measured (a) and reconstructed (b) SHG-FROG traces.
 (c) Retrieved spectrum (blue line), retrieved spectral phase (red line), and measured spectrum (gray line).
 (d) Retrieved temporal pulse profile (blue line) and temporal phase (red line). 
 }
 }
\end{figure}
%%%%%%%%%%%%%%%%%%%%%%%%%%%%%%%%%%%%%%%%%%

For dispersion management,  we replaced the CaF$_2$ prism pair 
between the second and third OPA stage in Ref.~\cite{malevich2016broadband} by Brewster Si plates, to change the sign of the chirp from negative to positive
to use Si bulk compressor with small TOD and a large throughput of 80 \%.
Here inside Si placed at the Brewster angle, the OPA beam is laterally expanded by a factor of $\approx 3.6$,
which lowers the pulse intensity and the B-integral of the pulse compressor.
For characterization of MIR pulses, we constructed a MIR SHG-FROG based on 2-mm-thick ZGP/200-$\mu$m-thick AGS crystals and
AOPDF-based scanning MIR spectrometer (MOZZA, Fastlite).
Figure 3(a) shows the first SHG-FROG measurement in the 5-10 $\mu$m region, which indicates
pulse compression to 99 fs (5.6 optical cycles) achieved within this simple GDD compensation scheme while FTL duration is 53 fs
and the difference comes from the large residual TOD of $6.3 \times 10^5$ fs$^3$ (Table 1).
Compensation TOD, which accumulates during the whole amplification processes due to its conserved sign through the two DFG processes
and where humidity in the air  \cite{mathar2007refractive} plays a crucial role at 5-7 $\mu$m (Table 1), will be presented in the next publication.
Here the energy stability of the OPA output and Ho:YAG RA output was  2.5\% rms and 1.2\% rms, respectively
and energy of parametric fluorescence measured when seed is blocked was less than 1 $\mu$J.
\vspace{-0.5cm}
%
%%%%%%%%%%%%%%%%%%%%%%%%%%%%%%%%%%%%%%%%%%
\renewcommand{\arraystretch}{1.2}
  \begin{table}[htb]
   \centering
      \caption{\bf Dispersion management for the MIR OPA.  }
  \begin{tabular}{l|cc} \hline 
       & $\phi^\mathrm{GDD}$ \scriptsize{($10^4$ fs$^2$)} & $\phi^\mathrm{TOD}$ \scriptsize{($10^4$ fs$^3$)}  \\ \hline 
  WLG at 1.5 $\mu$m & 0.6 & 0  \\  
  Material except air at 5.2 $\mu$m & 0.5 & 12.7  \\  
  Stretcher at 3.5 $\mu$m & 1.8 & 3.3  \\ 
  Compressor at 5.2 $\mu$m & 0.8 & 2.3  \\
  Air \scriptsize{(5.2 $\mu$m, 2 m, humidity 40\%)} & -0.3 & 34.5  \\ \hline
  Total (measured value) & -0.2 (-0.3) & 52.8 (63.1) \\ \hline
  \end{tabular}
  \end{table}
%%%%%%%%%%%%%%%%%%%%%%%%%%%%%%%%%%%%%%%%%%

Here it was crucial to avoid parasitic cascaded $\chi^2$ process \cite{desalvo1992self}, 
which easily elongate the pulse duration to a few picoseconds 
because of the exceptionally high $d_\mathrm{eff} \approx 80$ pm/V of the ZGP crystals, which is e.g., 40 times larger than 800 nm-pumped BBO OPA;
just optimizing pulse energy leads to strong phase modulation by this process
as shown in an SHG-FROG measurement and its retrieval (Fig.~4(a,b)), when one maximizes the pulse energy with a 8-mm-thick ZGP crystal.
Here retrieved phase shape can be expressed up to TOD, which can be compressed by a combination of a prism compressor and bulk,
while the spectrum supports the Fourier limited pulse duration of 42 fs (2.3 optical cycles). 
Therefore this phase modulation during parametric amplification opens a novel chirped pulse compression scheme.  
Here the relative sign between the chirp of the seed pulses and the phase obtained
from the cascaded $\chi2$ process was crucial for spectral broadening, 
while the latter can be both positive and negative according to the sign of phase mismatching $\Delta k$ \cite{desalvo1992self}.
%
%%%%%%%%%%%%%%%%%%%%%%%%%%%%%%%%%%%%%%%%%%
\begin{figure}[htbp]
 \centering
 \includegraphics[width=\linewidth]{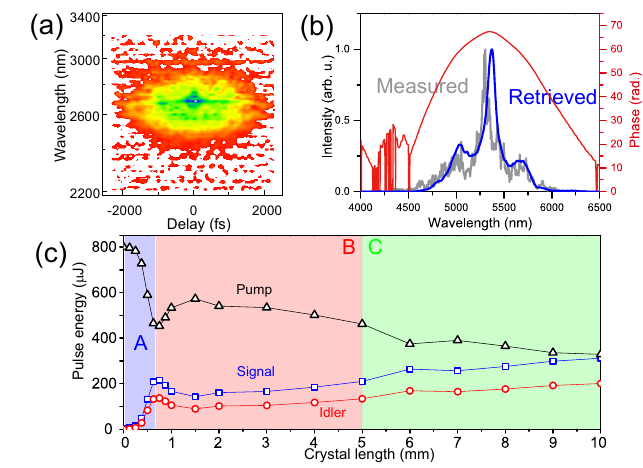}
 \caption{%
 \footnotesize{%
 FROG measurement for 5.2 $\mu$m pulses whose spectra strongly broadened by the cascaded $\chi^2$ process
 as indicated by the shoulder-like structure.  Novel approach for chirped pulse compression with narrowband seed.
 (a) Measured SHG-FROG traces.
 (b) Retrieved spectrum (blue line), retrieved spectral phase (red line), and measured spectrum (gray line).
 $\phi_\mathrm{GDD}=-2.6 \times 10^5$ fs$^2$ and
 $\phi_\mathrm{TOD}= 4.0 \times 10^6$ fs$^3$.
 (c) Calculated output pulse enegy for signal (blue square), idler (red circle), and pump (black triangle) as a function of ZGP crystals.
 }
 }
\end{figure}
%%%%%%%%%%%%%%%%%%%%%%%%%%%%%%%%%%%%%%%%%%
%

To understand the underlying physics, we performed numerical simulations based on Maxwell's equations in the slowly varying envelope approximation.
Figure 4(c) shows a calculated pulse energy of signal, idler, and pump as a function of
crystal length of ZGP.
Phase A in Fig.~4(c), which corresponds to Fig.~3's case, is normal parametric amplification region and here well known relations
$\phi^\mathrm{GDD}_\mathrm{i} \approx -\phi^\mathrm{GDD}_\mathrm{s }\approx -\phi^\mathrm{GDD}_\mathrm{seed}$ and
$\phi^\mathrm{TOD}_\mathrm{i} \approx \phi^\mathrm{TOD}_\mathrm{s} \approx \phi^\mathrm{GDD}_\mathrm{seed}$ are satisfied and the gain can be expressed as  $\exp (2 \Gamma L)/4$, where
$
\Gamma := {\sqrt{\omega_\mathrm{s}\omega_\mathrm{i}}|A_\mathrm{p0}| d_\mathrm{eff}}/{\sqrt{n_\mathrm{s}n_\mathrm{i}}c} \propto d_\mathrm{eff}
$,
which predicts exponentially ultrafast growing of seed pulses in ZGP.
Here $\phi^\mathrm{GDD}_\mu$ and $\phi^\mathrm{TOD}_\mu$, $\omega_\mu$, $n_\mu$ denote GDD, TOD values, angular frequency, refractive index for $\mu=$ s (signal), i (idler), and seed,
while $A_\mathrm{p0}$ and $c$ express incident electric field of pump and speed of light, respectively.
After the saturation of gain, whose crystal length was about 2 times longer than the calculated value mainly due to the degraded spacial profile of pump laser, the system enters Phase B.
This phase corresponds to Fig.~4 (a-b)'s case and the so-called back conversion from singal and idler to the pump started to take place and not only pulse duration of signal,
idler but also pump is elongated to a few picosecond, as can be seen in the Fig.~4(a).
Finally, in Phase C the pulse duration is elongated to about 5 ps and the broadened spectra of signal and idler started to be overlapped and high-frequency component of phase noise appears.
Pulse compression of these pulses becomes unrealistic while the output energy is larger than those of Phase A and B.
This phenomena is similar to the one described in Ref.~\cite{desalvo1992self} but the novel aspect in the present scheme is simultaneous realization of amplification (that is at small $|\Delta k|$) and spectral broadening,
which is realized only by the exceptionally large $d_\mathrm{eff}$.
%
%%%%%%%%%%%%%%%%%%%%%%%%%%%%%%%%%%%%%%%%%%
\begin{figure}[htbp]
 \centering
 \includegraphics[width=\linewidth]{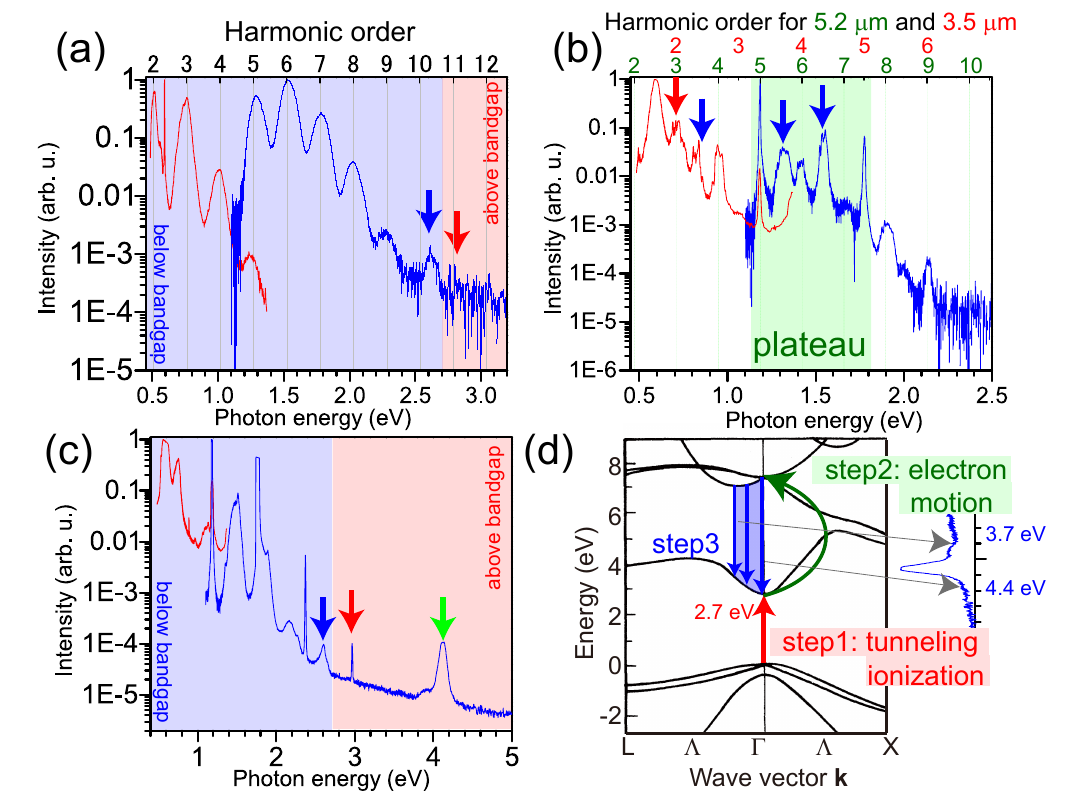}
 \caption{%
 \footnotesize{%
 (a) Observed HHG spectra driven by 5.2 $\mu$m pulses (40 $\mu$J, blue shifted during optimization) in a 1-mm-thick ZnSe crystal.
 Here the red and blue curves shows spectra measured by NIR and visible spectrometers(Ocean Optics, NIRQuest512 and HR4000) for (a-c). 
The peak around 2.6 eV is luminescence from exciton annihilation (blue arrows in (a,c)) and the observed 11th harmonic is above the bandgap of ZnSe (2.7 eV).
 (b) Observed two-color HHG spectra of 5.2 $\mu$m (10 $\mu$J) and 3.5 $\mu$m pulses (20 $\mu$J) in the same crystals.  
 As well as pure harmonics of  5.2 $\mu$m and 3.5 $\mu$m,  peaks by high-order sum- and difference- frequency mixing were observed (blue arrows).
 Notice CEP dependent fringe appears at e.g., 0.74 eV (red arrow).
 (c,d) Ultrafast intraband transition at 3.7 - 4.4 eV (green arrow), which corresponds to the process indicated by the blue arrows (step 3) in (d)
 (the band structure is from Ref.~\cite{chelikowsky1973calculated,Adachi1991}).
 Here the 5th harmonic of leaked pump pulses 2.96 eV above the bandgap (red arrow) is also observed.
 }
 }
\end{figure}
%%%%%%%%%%%%%%%%%%%%%%%%%%%%%%%%%%%%%%%%%%

%%%%%%%%%%%%%%%%%%%%%%%%%%%%%%%%%%%%%%%%%%%%%
%\section{application}
%%%%%%%%%%%%%%%%%%%%%%%%%%%%%%%%%%%%%%%%%%%%%

 In order to demonstrate applicability of generated MIR radiation in spectroscopy, nonlinear optics, and material science
 we have generated harmonics in bulk polycrystalline ZnSe plates from both 3.5 $\mu$m signal and 5.2 $\mu$m idler pulses having the same linear polarization and energies of 80 $\mu$J and 40 $\mu$J, respectively.
 The spectra generated in 1-mm thick ZnSe plate  are shown in Fig.~5. HHG were driven either only by idler pulses (Fig.~5(a)) or by both idler and signal pulses simultaneously (Fig.~5(b,c)).
 In the case of idler driver at  $\sim \!\!5$ $\mu$m, we have observed harmonics up to 11th order, which is above the bandgap of the ZnSe (2.7 eV).
 The peak in the vicinity of 2.6 eV, which is more clearly seen in Fig.~5(c), is attributed to the luminescence from excitonic states which are created by inter-band transitions (10-photon absorption in the case of $\sim \!\!5$-$\mu$m driver).
 In Fig.~5(b), where harmonics are generated by both, signal and idler pulses, high order sum- and difference-frequency mixing with a clear plateau is observed, which is an indication of the process taking place in a non-perturbative regime. Here some of the peaks have CEP dependent interference fringes because more than 2 paths create these signals 
 (e.g., for the peak at 0.72 eV $= 2 \hbar \omega_\mathrm{s}  = 3\hbar \omega_\mathrm{i}$).
 Finally, Fig.~5(c) shows a signal from ultrafast intra-band transition at $3.7$-$4.4$ eV.
 This asymmetric peak can be explained by the de-excitation process indicated by the blue arrows in Fig.~5(d) (taken from Ref.~\cite{chelikowsky1973calculated}).
 Considering that photon energy is above the bandgap, generally it should contain the information about the transient density of states as well as one about the band structure.

%%%%%%%%%%%%%%%%%%%%%%%%%%%%%%%%%%%%%%%%%%%%%
%\section{Conclusion}
%%%%%%%%%%%%%%%%%%%%%%%%%%%%%%%%%%%%%%%%%%%%%
 In conclusion, we have generated 99 fs, 5.2-$\mu$m, 40 $\mu$J pulses at a 1 kHz repetition rate.
 As the first application of this laser system, we generated high harmonics in bulk ZnSe crystals and observed 
 HHG above the bandgap, dense exciton generation after 10-photon absorption, high order sum- and difference-frequency generation,
 ultrafast transition in the conduction band, which reflects the structure of conduction band.  
%
%%%%%%%%%%%%%%%%%%%%%%%%%%%%%%%%%%%%
\section*{Funding Information}
%%%%%%%%%%%%%%%%%%%%%%%%%%%%%%%%%%%
Austrian Science Fund (FWF) (F4903-N23, SFB NextLite, M1884-N27, Samurai); 
European Research Council (ERC) (620316, Luminos); 
\"{O}sterreichische Forschungsf\"orderungsgesellschaft (FFG) (E! 6698, Mirandus);
Bundesministerium f\"ur Bildung und Forschung (BMBF) (E! 6698, Mirandus).

\bibliography{optlett4}

\begin{thebibliography}{27}%
\makeatletter
\providecommand \@ifxundefined [1]{%
 \@ifx{#1\undefined}
}%
\providecommand \@ifnum [1]{%
 \ifnum #1\expandafter \@firstoftwo
 \else \expandafter \@secondoftwo
 \fi
}%
\providecommand \@ifx [1]{%
 \ifx #1\expandafter \@firstoftwo
 \else \expandafter \@secondoftwo
 \fi
}%
\providecommand \natexlab [1]{#1}%
\providecommand \enquote  [1]{``#1''}%
\providecommand \bibnamefont  [1]{#1}%
\providecommand \bibfnamefont [1]{#1}%
\providecommand \citenamefont [1]{#1}%
\providecommand \href@noop [0]{\@secondoftwo}%
\providecommand \href [0]{\begingroup \@sanitize@url \@href}%
\providecommand \@href[1]{\@@startlink{#1}\@@href}%
\providecommand \@@href[1]{\endgroup#1\@@endlink}%
\providecommand \@sanitize@url [0]{\catcode `\\12\catcode `\$12\catcode
  `\&12\catcode `\#12\catcode `\^12\catcode `\_12\catcode `\%12\relax}%
\providecommand \@@startlink[1]{}%
\providecommand \@@endlink[0]{}%
\providecommand \url  [0]{\begingroup\@sanitize@url \@url }%
\providecommand \@url [1]{\endgroup\@href {#1}{\urlprefix }}%
\providecommand \urlprefix  [0]{URL }%
\providecommand \Eprint [0]{\href }%
\providecommand \doibase [0]{http://dx.doi.org/}%
\providecommand \selectlanguage [0]{\@gobble}%
\providecommand \bibinfo  [0]{\@secondoftwo}%
\providecommand \bibfield  [0]{\@secondoftwo}%
\providecommand \translation [1]{[#1]}%
\providecommand \BibitemOpen [0]{}%
\providecommand \bibitemStop [0]{}%
\providecommand \bibitemNoStop [0]{.\EOS\space}%
\providecommand \EOS [0]{\spacefactor3000\relax}%
\providecommand \BibitemShut  [1]{\csname bibitem#1\endcsname}%
\let\auto@bib@innerbib\@empty
%</preamble>
\bibitem [{\citenamefont {Malevich}\ \emph {et~al.}(2016)\citenamefont
  {Malevich}, \citenamefont {Kanai}, \citenamefont {Hoogland}, \citenamefont
  {Holzwarth}, \citenamefont {Baltu{\v{s}}ka},\ and\ \citenamefont
  {Pug{\v{z}}lys}}]{malevich2016broadband}%
  \BibitemOpen
  \bibfield  {author} {\bibinfo {author} {\bibfnamefont {P.}~\bibnamefont
  {Malevich}}, \bibinfo {author} {\bibfnamefont {T.}~\bibnamefont {Kanai}},
  \bibinfo {author} {\bibfnamefont {H.}~\bibnamefont {Hoogland}}, \bibinfo
  {author} {\bibfnamefont {R.}~\bibnamefont {Holzwarth}}, \bibinfo {author}
  {\bibfnamefont {A.}~\bibnamefont {Baltu{\v{s}}ka}}, \ and\ \bibinfo {author}
  {\bibfnamefont {A.}~\bibnamefont {Pug{\v{z}}lys}},\ }\href@noop {} {\bibfield
   {journal} {\bibinfo  {journal} {Opt. Lett.}\ }\textbf {\bibinfo {volume}
  {41}},\ \bibinfo {pages} {930} (\bibinfo {year} {2016})}\BibitemShut
  {NoStop}%
\bibitem [{\citenamefont {Sanchez}\ \emph {et~al.}(2016)\citenamefont
  {Sanchez}, \citenamefont {Hemmer}, \citenamefont {Baudisch}, \citenamefont
  {Cousin}, \citenamefont {Zawilski}, \citenamefont {Schunemann}, \citenamefont
  {Chalus}, \citenamefont {Simon-Boisson},\ and\ \citenamefont
  {Biegert}}]{sanchez20167}%
  \BibitemOpen
  \bibfield  {author} {\bibinfo {author} {\bibfnamefont {D.}~\bibnamefont
  {Sanchez}}, \bibinfo {author} {\bibfnamefont {M.}~\bibnamefont {Hemmer}},
  \bibinfo {author} {\bibfnamefont {M.}~\bibnamefont {Baudisch}}, \bibinfo
  {author} {\bibfnamefont {S.}~\bibnamefont {Cousin}}, \bibinfo {author}
  {\bibfnamefont {K.}~\bibnamefont {Zawilski}}, \bibinfo {author}
  {\bibfnamefont {P.}~\bibnamefont {Schunemann}}, \bibinfo {author}
  {\bibfnamefont {O.}~\bibnamefont {Chalus}}, \bibinfo {author} {\bibfnamefont
  {C.}~\bibnamefont {Simon-Boisson}}, \ and\ \bibinfo {author} {\bibfnamefont
  {J.}~\bibnamefont {Biegert}},\ }\href@noop {} {\bibfield  {journal} {\bibinfo
   {journal} {Optica}\ }\textbf {\bibinfo {volume} {3}},\ \bibinfo {pages}
  {147} (\bibinfo {year} {2016})}\BibitemShut {NoStop}%
\bibitem [{\citenamefont {Andriukaitis}\ \emph {et~al.}(2011)\citenamefont
  {Andriukaitis}, \citenamefont {Bal{\v{c}}i{\=u}nas}, \citenamefont
  {Ali{\v{s}}auskas}, \citenamefont {Pug{\v{z}}lys}, \citenamefont
  {Baltu{\v{s}}ka}, \citenamefont {Popmintchev}, \citenamefont {Chen},
  \citenamefont {Murnane},\ and\ \citenamefont {Kapteyn}}]{Andriukaitis2011}%
  \BibitemOpen
  \bibfield  {author} {\bibinfo {author} {\bibfnamefont {G.}~\bibnamefont
  {Andriukaitis}}, \bibinfo {author} {\bibfnamefont {T.}~\bibnamefont
  {Bal{\v{c}}i{\=u}nas}}, \bibinfo {author} {\bibfnamefont {S.}~\bibnamefont
  {Ali{\v{s}}auskas}}, \bibinfo {author} {\bibfnamefont {A.}~\bibnamefont
  {Pug{\v{z}}lys}}, \bibinfo {author} {\bibfnamefont {A.}~\bibnamefont
  {Baltu{\v{s}}ka}}, \bibinfo {author} {\bibfnamefont {T.}~\bibnamefont
  {Popmintchev}}, \bibinfo {author} {\bibfnamefont {M.-C.}\ \bibnamefont
  {Chen}}, \bibinfo {author} {\bibfnamefont {M.~M.}\ \bibnamefont {Murnane}}, \
  and\ \bibinfo {author} {\bibfnamefont {H.~C.}\ \bibnamefont {Kapteyn}},\
  }\href {\doibase 10.1364/OL.36.002755} {\bibfield  {journal} {\bibinfo
  {journal} {Opt. Lett.}\ }\textbf {\bibinfo {volume} {36}},\ \bibinfo {pages}
  {2755} (\bibinfo {year} {2011})}\BibitemShut {NoStop}%
\bibitem [{\citenamefont {Popmintchev}\ \emph {et~al.}(2012)\citenamefont
  {Popmintchev}, \citenamefont {Chen}, \citenamefont {Popmintchev},
  \citenamefont {Arpin}, \citenamefont {Brown}, \citenamefont
  {Ali{\v{s}}auskas}, \citenamefont {Andriukaitis}, \citenamefont
  {Bal{\v{c}}i{\=u}nas}, \citenamefont {M{\"u}cke}, \citenamefont {Pugzlys}
  \emph {et~al.}}]{popmintchev2012bright}%
  \BibitemOpen
  \bibfield  {author} {\bibinfo {author} {\bibfnamefont {T.}~\bibnamefont
  {Popmintchev}}, \bibinfo {author} {\bibfnamefont {M.-C.}\ \bibnamefont
  {Chen}}, \bibinfo {author} {\bibfnamefont {D.}~\bibnamefont {Popmintchev}},
  \bibinfo {author} {\bibfnamefont {P.}~\bibnamefont {Arpin}}, \bibinfo
  {author} {\bibfnamefont {S.}~\bibnamefont {Brown}}, \bibinfo {author}
  {\bibfnamefont {S.}~\bibnamefont {Ali{\v{s}}auskas}}, \bibinfo {author}
  {\bibfnamefont {G.}~\bibnamefont {Andriukaitis}}, \bibinfo {author}
  {\bibfnamefont {T.}~\bibnamefont {Bal{\v{c}}i{\=u}nas}}, \bibinfo {author}
  {\bibfnamefont {O.~D.}\ \bibnamefont {M{\"u}cke}}, \bibinfo {author}
  {\bibfnamefont {A.}~\bibnamefont {Pugzlys}},  \emph {et~al.},\ }\href@noop {}
  {\bibfield  {journal} {\bibinfo  {journal} {Science}\ }\textbf {\bibinfo
  {volume} {336}},\ \bibinfo {pages} {1287} (\bibinfo {year}
  {2012})}\BibitemShut {NoStop}%
\bibitem [{\citenamefont {Weisshaupt}\ \emph {et~al.}(2014)\citenamefont
  {Weisshaupt}, \citenamefont {Juv{\'e}}, \citenamefont {Holtz}, \citenamefont
  {Ku}, \citenamefont {Woerner}, \citenamefont {Elsaesser}, \citenamefont
  {Ali{\v{s}}auskas}, \citenamefont {Pug{\v{z}}lys},\ and\ \citenamefont
  {Baltu{\v{s}}ka}}]{weisshaupt2014high}%
  \BibitemOpen
  \bibfield  {author} {\bibinfo {author} {\bibfnamefont {J.}~\bibnamefont
  {Weisshaupt}}, \bibinfo {author} {\bibfnamefont {V.}~\bibnamefont
  {Juv{\'e}}}, \bibinfo {author} {\bibfnamefont {M.}~\bibnamefont {Holtz}},
  \bibinfo {author} {\bibfnamefont {S.}~\bibnamefont {Ku}}, \bibinfo {author}
  {\bibfnamefont {M.}~\bibnamefont {Woerner}}, \bibinfo {author} {\bibfnamefont
  {T.}~\bibnamefont {Elsaesser}}, \bibinfo {author} {\bibfnamefont
  {S.}~\bibnamefont {Ali{\v{s}}auskas}}, \bibinfo {author} {\bibfnamefont
  {A.}~\bibnamefont {Pug{\v{z}}lys}}, \ and\ \bibinfo {author} {\bibfnamefont
  {A.}~\bibnamefont {Baltu{\v{s}}ka}},\ }\href@noop {} {\bibfield  {journal}
  {\bibinfo  {journal} {Nat. Photonics}\ }\textbf {\bibinfo {volume} {8}},\
  \bibinfo {pages} {927} (\bibinfo {year} {2014})}\BibitemShut {NoStop}%
\bibitem [{\citenamefont {Mitrofanov}\ \emph {et~al.}(2015)\citenamefont
  {Mitrofanov}, \citenamefont {Voronin}, \citenamefont {Sidorov-Biryukov},
  \citenamefont {Pug{\v{z}}lys}, \citenamefont {Stepanov}, \citenamefont
  {Andriukaitis}, \citenamefont {Fl{\"o}ry}, \citenamefont {Ali{\v{s}}auskas},
  \citenamefont {Fedotov}, \citenamefont {Baltu{\v{s}}ka},\ and\ \citenamefont
  {Zheltikov}}]{mitrofanov2015mid}%
  \BibitemOpen
  \bibfield  {author} {\bibinfo {author} {\bibfnamefont {A.}~\bibnamefont
  {Mitrofanov}}, \bibinfo {author} {\bibfnamefont {A.}~\bibnamefont {Voronin}},
  \bibinfo {author} {\bibfnamefont {D.}~\bibnamefont {Sidorov-Biryukov}},
  \bibinfo {author} {\bibfnamefont {A.}~\bibnamefont {Pug{\v{z}}lys}}, \bibinfo
  {author} {\bibfnamefont {E.}~\bibnamefont {Stepanov}}, \bibinfo {author}
  {\bibfnamefont {G.}~\bibnamefont {Andriukaitis}}, \bibinfo {author}
  {\bibfnamefont {T.}~\bibnamefont {Fl{\"o}ry}}, \bibinfo {author}
  {\bibfnamefont {S.}~\bibnamefont {Ali{\v{s}}auskas}}, \bibinfo {author}
  {\bibfnamefont {A.}~\bibnamefont {Fedotov}}, \bibinfo {author} {\bibfnamefont
  {A.}~\bibnamefont {Baltu{\v{s}}ka}}, \ and\ \bibinfo {author} {\bibfnamefont
  {A.}~\bibnamefont {Zheltikov}},\ }\href@noop {} {\bibfield  {journal}
  {\bibinfo  {journal} {Sci. Rep.}\ }\textbf {\bibinfo {volume} {5}} (\bibinfo
  {year} {2015})}\BibitemShut {NoStop}%
\bibitem [{\citenamefont {Nomura}\ \emph {et~al.}(2013)\citenamefont {Nomura},
  \citenamefont {Shirai},\ and\ \citenamefont {Fuji}}]{nomura2013frequency}%
  \BibitemOpen
  \bibfield  {author} {\bibinfo {author} {\bibfnamefont {Y.}~\bibnamefont
  {Nomura}}, \bibinfo {author} {\bibfnamefont {H.}~\bibnamefont {Shirai}}, \
  and\ \bibinfo {author} {\bibfnamefont {T.}~\bibnamefont {Fuji}},\ }\href@noop
  {} {\bibfield  {journal} {\bibinfo  {journal} {Nat. Commun.}\ }\textbf
  {\bibinfo {volume} {4}} (\bibinfo {year} {2013})}\BibitemShut {NoStop}%
\bibitem [{\citenamefont {Liang}\ \emph {et~al.}(2016)\citenamefont {Liang},
  \citenamefont {Krogen}, \citenamefont {Zawilski}, \citenamefont {Schunemann},
  \citenamefont {Lang}, \citenamefont {Morgner}, \citenamefont {K{\"a}rtner},
  \citenamefont {Moses},\ and\ \citenamefont {Hong}}]{liang2016octave}%
  \BibitemOpen
  \bibfield  {author} {\bibinfo {author} {\bibfnamefont {H.~K.}\ \bibnamefont
  {Liang}}, \bibinfo {author} {\bibfnamefont {P.}~\bibnamefont {Krogen}},
  \bibinfo {author} {\bibfnamefont {K.}~\bibnamefont {Zawilski}}, \bibinfo
  {author} {\bibfnamefont {P.~G.}\ \bibnamefont {Schunemann}}, \bibinfo
  {author} {\bibfnamefont {T.}~\bibnamefont {Lang}}, \bibinfo {author}
  {\bibfnamefont {U.}~\bibnamefont {Morgner}}, \bibinfo {author} {\bibfnamefont
  {F.}~\bibnamefont {K{\"a}rtner}}, \bibinfo {author} {\bibfnamefont
  {J.}~\bibnamefont {Moses}}, \ and\ \bibinfo {author} {\bibfnamefont {K.-H.}\
  \bibnamefont {Hong}},\ }in\ \href@noop {} {\emph {\bibinfo {booktitle}
  {Mid-Infrared Coherent Sources}}}\ (\bibinfo {organization} {Optical Society
  of America},\ \bibinfo {year} {2016})\ p.\ \bibinfo {pages}
  {MS4C{.}1}\BibitemShut {NoStop}%
\bibitem [{\citenamefont {Wandel}\ \emph {et~al.}(2016)\citenamefont {Wandel},
  \citenamefont {Lin}, \citenamefont {Yin}, \citenamefont {Xu},\ and\
  \citenamefont {Jovanovic}}]{wandel2016parametric}%
  \BibitemOpen
  \bibfield  {author} {\bibinfo {author} {\bibfnamefont {S.}~\bibnamefont
  {Wandel}}, \bibinfo {author} {\bibfnamefont {M.-W.}\ \bibnamefont {Lin}},
  \bibinfo {author} {\bibfnamefont {Y.}~\bibnamefont {Yin}}, \bibinfo {author}
  {\bibfnamefont {G.}~\bibnamefont {Xu}}, \ and\ \bibinfo {author}
  {\bibfnamefont {I.}~\bibnamefont {Jovanovic}},\ }\href@noop {} {\bibfield
  {journal} {\bibinfo  {journal} {Opt. Express}\ }\textbf {\bibinfo {volume}
  {24}},\ \bibinfo {pages} {5287} (\bibinfo {year} {2016})}\BibitemShut
  {NoStop}%
\bibitem [{\citenamefont {Golubovic}\ and\ \citenamefont
  {Reed}(1998)}]{golubovic1998all}%
  \BibitemOpen
  \bibfield  {author} {\bibinfo {author} {\bibfnamefont {B.}~\bibnamefont
  {Golubovic}}\ and\ \bibinfo {author} {\bibfnamefont {M.}~\bibnamefont
  {Reed}},\ }\href@noop {} {\bibfield  {journal} {\bibinfo  {journal} {Opt.
  Lett.}\ }\textbf {\bibinfo {volume} {23}},\ \bibinfo {pages} {1760} (\bibinfo
  {year} {1998})}\BibitemShut {NoStop}%
\bibitem [{\citenamefont {Lanin}\ \emph {et~al.}(2014)\citenamefont {Lanin},
  \citenamefont {Fedotov},\ and\ \citenamefont {Zheltikov}}]{Lanin2014}%
  \BibitemOpen
  \bibfield  {author} {\bibinfo {author} {\bibfnamefont {A.~A.}\ \bibnamefont
  {Lanin}}, \bibinfo {author} {\bibfnamefont {A.~B.}\ \bibnamefont {Fedotov}},
  \ and\ \bibinfo {author} {\bibfnamefont {A.~M.}\ \bibnamefont {Zheltikov}},\
  }\href {\doibase 10.1364/JOSAB.31.001901} {\bibfield  {journal} {\bibinfo
  {journal} {J. Opt. Soc. Am. B}\ }\textbf {\bibinfo {volume} {31}},\ \bibinfo
  {pages} {1901} (\bibinfo {year} {2014})}\BibitemShut {NoStop}%
\bibitem [{\citenamefont {Petrov}\ \emph
  {et~al.}(2004{\natexlab{a}})\citenamefont {Petrov}, \citenamefont
  {Yelisseyev}, \citenamefont {Isaenko}, \citenamefont {Lobanov}, \citenamefont
  {Titov},\ and\ \citenamefont {Zondy}}]{petrov2004second}%
  \BibitemOpen
  \bibfield  {author} {\bibinfo {author} {\bibfnamefont {V.}~\bibnamefont
  {Petrov}}, \bibinfo {author} {\bibfnamefont {A.}~\bibnamefont {Yelisseyev}},
  \bibinfo {author} {\bibfnamefont {L.}~\bibnamefont {Isaenko}}, \bibinfo
  {author} {\bibfnamefont {S.}~\bibnamefont {Lobanov}}, \bibinfo {author}
  {\bibfnamefont {A.}~\bibnamefont {Titov}}, \ and\ \bibinfo {author}
  {\bibfnamefont {J.-J.}\ \bibnamefont {Zondy}},\ }\href@noop {} {\bibfield
  {journal} {\bibinfo  {journal} {Appl. Phys. B}\ }\textbf {\bibinfo {volume}
  {78}},\ \bibinfo {pages} {543} (\bibinfo {year}
  {2004}{\natexlab{a}})}\BibitemShut {NoStop}%
\bibitem [{\citenamefont {Rotermund}\ \emph {et~al.}(2001)\citenamefont
  {Rotermund}, \citenamefont {Petrov}, \citenamefont {Noack}, \citenamefont
  {Isaenko}, \citenamefont {Yelisseyev},\ and\ \citenamefont
  {Lobanov}}]{rotermund2001optical}%
  \BibitemOpen
  \bibfield  {author} {\bibinfo {author} {\bibfnamefont {F.}~\bibnamefont
  {Rotermund}}, \bibinfo {author} {\bibfnamefont {V.}~\bibnamefont {Petrov}},
  \bibinfo {author} {\bibfnamefont {F.}~\bibnamefont {Noack}}, \bibinfo
  {author} {\bibfnamefont {L.}~\bibnamefont {Isaenko}}, \bibinfo {author}
  {\bibfnamefont {A.}~\bibnamefont {Yelisseyev}}, \ and\ \bibinfo {author}
  {\bibfnamefont {S.}~\bibnamefont {Lobanov}},\ }\href@noop {} {\bibfield
  {journal} {\bibinfo  {journal} {Appl. Phys. Lett.}\ }\textbf {\bibinfo
  {volume} {78}},\ \bibinfo {pages} {2623} (\bibinfo {year}
  {2001})}\BibitemShut {NoStop}%
\bibitem [{\citenamefont {Petrov}\ \emph
  {et~al.}(2004{\natexlab{b}})\citenamefont {Petrov}, \citenamefont {Badikov},
  \citenamefont {Panyutin}, \citenamefont {Shevyrdyaeva}, \citenamefont
  {Sheina},\ and\ \citenamefont {Rotermund}}]{petrov2004mid}%
  \BibitemOpen
  \bibfield  {author} {\bibinfo {author} {\bibfnamefont {V.}~\bibnamefont
  {Petrov}}, \bibinfo {author} {\bibfnamefont {V.}~\bibnamefont {Badikov}},
  \bibinfo {author} {\bibfnamefont {V.}~\bibnamefont {Panyutin}}, \bibinfo
  {author} {\bibfnamefont {G.}~\bibnamefont {Shevyrdyaeva}}, \bibinfo {author}
  {\bibfnamefont {S.}~\bibnamefont {Sheina}}, \ and\ \bibinfo {author}
  {\bibfnamefont {F.}~\bibnamefont {Rotermund}},\ }\href@noop {} {\bibfield
  {journal} {\bibinfo  {journal} {Opt. Commun.}\ }\textbf {\bibinfo {volume}
  {235}},\ \bibinfo {pages} {219} (\bibinfo {year}
  {2004}{\natexlab{b}})}\BibitemShut {NoStop}%
\bibitem [{\citenamefont {Kaindl}\ \emph {et~al.}(1999)\citenamefont {Kaindl},
  \citenamefont {Eickemeyer}, \citenamefont {Woerner},\ and\ \citenamefont
  {Elsaesser}}]{kaindl1999broadband}%
  \BibitemOpen
  \bibfield  {author} {\bibinfo {author} {\bibfnamefont {R.}~\bibnamefont
  {Kaindl}}, \bibinfo {author} {\bibfnamefont {F.}~\bibnamefont {Eickemeyer}},
  \bibinfo {author} {\bibfnamefont {M.}~\bibnamefont {Woerner}}, \ and\
  \bibinfo {author} {\bibfnamefont {T.}~\bibnamefont {Elsaesser}},\ }\href@noop
  {} {\bibfield  {journal} {\bibinfo  {journal} {Appl. Phys. Lett.}\ }\textbf
  {\bibinfo {volume} {75}},\ \bibinfo {pages} {1060} (\bibinfo {year}
  {1999})}\BibitemShut {NoStop}%
\bibitem [{\citenamefont {DeSalvo}\ \emph {et~al.}(1992)\citenamefont
  {DeSalvo}, \citenamefont {Vanherzeele}, \citenamefont {Hagan}, \citenamefont
  {Sheik-Bahae}, \citenamefont {Stegeman},\ and\ \citenamefont
  {Van~Stryland}}]{desalvo1992self}%
  \BibitemOpen
  \bibfield  {author} {\bibinfo {author} {\bibfnamefont {R.}~\bibnamefont
  {DeSalvo}}, \bibinfo {author} {\bibfnamefont {H.}~\bibnamefont
  {Vanherzeele}}, \bibinfo {author} {\bibfnamefont {D.}~\bibnamefont {Hagan}},
  \bibinfo {author} {\bibfnamefont {M.}~\bibnamefont {Sheik-Bahae}}, \bibinfo
  {author} {\bibfnamefont {G.}~\bibnamefont {Stegeman}}, \ and\ \bibinfo
  {author} {\bibfnamefont {E.}~\bibnamefont {Van~Stryland}},\ }\href@noop {}
  {\bibfield  {journal} {\bibinfo  {journal} {Opt. Lett.}\ }\textbf {\bibinfo
  {volume} {17}},\ \bibinfo {pages} {28} (\bibinfo {year} {1992})}\BibitemShut
  {NoStop}%
\bibitem [{\citenamefont {Luu}\ \emph {et~al.}(2015)\citenamefont {Luu},
  \citenamefont {Garg}, \citenamefont {Kruchinin}, \citenamefont {Moulet},
  \citenamefont {Hassan},\ and\ \citenamefont {Goulielmakis}}]{luu2015extreme}%
  \BibitemOpen
  \bibfield  {author} {\bibinfo {author} {\bibfnamefont {T.~T.}\ \bibnamefont
  {Luu}}, \bibinfo {author} {\bibfnamefont {M.}~\bibnamefont {Garg}}, \bibinfo
  {author} {\bibfnamefont {S.~Y.}\ \bibnamefont {Kruchinin}}, \bibinfo {author}
  {\bibfnamefont {A.}~\bibnamefont {Moulet}}, \bibinfo {author} {\bibfnamefont
  {M.~T.}\ \bibnamefont {Hassan}}, \ and\ \bibinfo {author} {\bibfnamefont
  {E.}~\bibnamefont {Goulielmakis}},\ }\href@noop {} {\bibfield  {journal}
  {\bibinfo  {journal} {Nature}\ }\textbf {\bibinfo {volume} {521}},\ \bibinfo
  {pages} {498} (\bibinfo {year} {2015})}\BibitemShut {NoStop}%
\bibitem [{\citenamefont {Ghimire}\ \emph {et~al.}(2011)\citenamefont
  {Ghimire}, \citenamefont {DiChiara}, \citenamefont {Sistrunk}, \citenamefont
  {Agostini}, \citenamefont {DiMauro},\ and\ \citenamefont
  {Reis}}]{ghimire2011observation}%
  \BibitemOpen
  \bibfield  {author} {\bibinfo {author} {\bibfnamefont {S.}~\bibnamefont
  {Ghimire}}, \bibinfo {author} {\bibfnamefont {A.~D.}\ \bibnamefont
  {DiChiara}}, \bibinfo {author} {\bibfnamefont {E.}~\bibnamefont {Sistrunk}},
  \bibinfo {author} {\bibfnamefont {P.}~\bibnamefont {Agostini}}, \bibinfo
  {author} {\bibfnamefont {L.~F.}\ \bibnamefont {DiMauro}}, \ and\ \bibinfo
  {author} {\bibfnamefont {D.~A.}\ \bibnamefont {Reis}},\ }\href@noop {}
  {\bibfield  {journal} {\bibinfo  {journal} {Nat. Phys.}\ }\textbf {\bibinfo
  {volume} {7}},\ \bibinfo {pages} {138} (\bibinfo {year} {2011})}\BibitemShut
  {NoStop}%
\bibitem [{\citenamefont {Vampa}\ \emph {et~al.}(2015)\citenamefont {Vampa},
  \citenamefont {Hammond}, \citenamefont {Thir{\'e}}, \citenamefont {Schmidt},
  \citenamefont {L{\'e}gar{\'e}}, \citenamefont {McDonald}, \citenamefont
  {Brabec},\ and\ \citenamefont {Corkum}}]{vampa2015linking}%
  \BibitemOpen
  \bibfield  {author} {\bibinfo {author} {\bibfnamefont {G.}~\bibnamefont
  {Vampa}}, \bibinfo {author} {\bibfnamefont {T.}~\bibnamefont {Hammond}},
  \bibinfo {author} {\bibfnamefont {N.}~\bibnamefont {Thir{\'e}}}, \bibinfo
  {author} {\bibfnamefont {B.}~\bibnamefont {Schmidt}}, \bibinfo {author}
  {\bibfnamefont {F.}~\bibnamefont {L{\'e}gar{\'e}}}, \bibinfo {author}
  {\bibfnamefont {C.}~\bibnamefont {McDonald}}, \bibinfo {author}
  {\bibfnamefont {T.}~\bibnamefont {Brabec}}, \ and\ \bibinfo {author}
  {\bibfnamefont {P.}~\bibnamefont {Corkum}},\ }\href@noop {} {\bibfield
  {journal} {\bibinfo  {journal} {Nature}\ }\textbf {\bibinfo {volume} {522}},\
  \bibinfo {pages} {462} (\bibinfo {year} {2015})}\BibitemShut {NoStop}%
\bibitem [{\citenamefont {Chin}\ \emph {et~al.}(2001)\citenamefont {Chin},
  \citenamefont {Calder\'{o}n},\ and\ \citenamefont {Kono}}]{Chin2001}%
  \BibitemOpen
  \bibfield  {author} {\bibinfo {author} {\bibfnamefont {A.~H.}\ \bibnamefont
  {Chin}}, \bibinfo {author} {\bibfnamefont {O.~G.}\ \bibnamefont
  {Calder\'{o}n}}, \ and\ \bibinfo {author} {\bibfnamefont {J.}~\bibnamefont
  {Kono}},\ }\href {\doibase 10.1103/PhysRevLett.86.3292} {\bibfield  {journal}
  {\bibinfo  {journal} {Phys. Rev. Lett.}\ }\textbf {\bibinfo {volume} {86}},\
  \bibinfo {pages} {3292} (\bibinfo {year} {2001})}\BibitemShut {NoStop}%
\bibitem [{\citenamefont {Haessler}\ \emph {et~al.}(2015)\citenamefont
  {Haessler}, \citenamefont {Bal{\v{c}}i{\=u}nas}, \citenamefont {Fan},
  \citenamefont {Chipperfield},\ and\ \citenamefont
  {Baltu{\v{s}}ka}}]{haessler2015enhanced}%
  \BibitemOpen
  \bibfield  {author} {\bibinfo {author} {\bibfnamefont {S.}~\bibnamefont
  {Haessler}}, \bibinfo {author} {\bibfnamefont {T.}~\bibnamefont
  {Bal{\v{c}}i{\=u}nas}}, \bibinfo {author} {\bibfnamefont {G.}~\bibnamefont
  {Fan}}, \bibinfo {author} {\bibfnamefont {L.~E.}\ \bibnamefont
  {Chipperfield}}, \ and\ \bibinfo {author} {\bibfnamefont {A.}~\bibnamefont
  {Baltu{\v{s}}ka}},\ }\href@noop {} {\bibfield  {journal} {\bibinfo  {journal}
  {Sci. Rep.}\ }\textbf {\bibinfo {volume} {5}} (\bibinfo {year}
  {2015})}\BibitemShut {NoStop}%
\bibitem [{\citenamefont {Hoogland}\ \emph {et~al.}(2014)\citenamefont
  {Hoogland}, \citenamefont {Wittek}, \citenamefont {H{\"a}nsel}, \citenamefont
  {Stark},\ and\ \citenamefont {Holzwarth}}]{hoogland2014fiber}%
  \BibitemOpen
  \bibfield  {author} {\bibinfo {author} {\bibfnamefont {H.}~\bibnamefont
  {Hoogland}}, \bibinfo {author} {\bibfnamefont {S.}~\bibnamefont {Wittek}},
  \bibinfo {author} {\bibfnamefont {W.}~\bibnamefont {H{\"a}nsel}}, \bibinfo
  {author} {\bibfnamefont {S.}~\bibnamefont {Stark}}, \ and\ \bibinfo {author}
  {\bibfnamefont {R.}~\bibnamefont {Holzwarth}},\ }\href@noop {} {\bibfield
  {journal} {\bibinfo  {journal} {Opt. Lett.}\ }\textbf {\bibinfo {volume}
  {39}},\ \bibinfo {pages} {6735} (\bibinfo {year} {2014})}\BibitemShut
  {NoStop}%
\bibitem [{\citenamefont {Malevich}\ \emph {et~al.}(2013)\citenamefont
  {Malevich}, \citenamefont {Andriukaitis}, \citenamefont {Fl{\"o}ry},
  \citenamefont {Verhoef}, \citenamefont {Fern{\'a}ndez}, \citenamefont
  {Ali{\v{s}}auskas}, \citenamefont {Pug{\v{z}}lys}, \citenamefont
  {Baltu{\v{s}}ka}, \citenamefont {Tan}, \citenamefont {Chua} \emph
  {et~al.}}]{malevich2013high}%
  \BibitemOpen
  \bibfield  {author} {\bibinfo {author} {\bibfnamefont {P.}~\bibnamefont
  {Malevich}}, \bibinfo {author} {\bibfnamefont {G.}~\bibnamefont
  {Andriukaitis}}, \bibinfo {author} {\bibfnamefont {T.}~\bibnamefont
  {Fl{\"o}ry}}, \bibinfo {author} {\bibfnamefont {A.}~\bibnamefont {Verhoef}},
  \bibinfo {author} {\bibfnamefont {A.}~\bibnamefont {Fern{\'a}ndez}}, \bibinfo
  {author} {\bibfnamefont {S.}~\bibnamefont {Ali{\v{s}}auskas}}, \bibinfo
  {author} {\bibfnamefont {A.}~\bibnamefont {Pug{\v{z}}lys}}, \bibinfo {author}
  {\bibfnamefont {A.}~\bibnamefont {Baltu{\v{s}}ka}}, \bibinfo {author}
  {\bibfnamefont {L.}~\bibnamefont {Tan}}, \bibinfo {author} {\bibfnamefont
  {C.}~\bibnamefont {Chua}},  \emph {et~al.},\ }\href@noop {} {\bibfield
  {journal} {\bibinfo  {journal} {Opt. Lett.}\ }\textbf {\bibinfo {volume}
  {38}},\ \bibinfo {pages} {2746} (\bibinfo {year} {2013})}\BibitemShut
  {NoStop}%
\bibitem [{\citenamefont {Baltu{\v{s}}ka}\ \emph {et~al.}(2002)\citenamefont
  {Baltu{\v{s}}ka}, \citenamefont {Fuji},\ and\ \citenamefont
  {Kobayashi}}]{baltuvska2002controlling}%
  \BibitemOpen
  \bibfield  {author} {\bibinfo {author} {\bibfnamefont {A.}~\bibnamefont
  {Baltu{\v{s}}ka}}, \bibinfo {author} {\bibfnamefont {T.}~\bibnamefont
  {Fuji}}, \ and\ \bibinfo {author} {\bibfnamefont {T.}~\bibnamefont
  {Kobayashi}},\ }\href@noop {} {\bibfield  {journal} {\bibinfo  {journal}
  {Phys. Rev. Lett.}\ }\textbf {\bibinfo {volume} {88}},\ \bibinfo {pages}
  {133901} (\bibinfo {year} {2002})}\BibitemShut {NoStop}%
\bibitem [{\citenamefont {Mathar}(2007)}]{mathar2007refractive}%
  \BibitemOpen
  \bibfield  {author} {\bibinfo {author} {\bibfnamefont {R.~J.}\ \bibnamefont
  {Mathar}},\ }\href@noop {} {\bibfield  {journal} {\bibinfo  {journal} {J.
  Opt. A: Pure Appl. Opt.}\ }\textbf {\bibinfo {volume} {9}},\ \bibinfo {pages}
  {470} (\bibinfo {year} {2007})}\BibitemShut {NoStop}%
\bibitem [{\citenamefont {Chelikowsky}\ \emph {et~al.}(1973)\citenamefont
  {Chelikowsky}, \citenamefont {Chadi},\ and\ \citenamefont
  {Cohen}}]{chelikowsky1973calculated}%
  \BibitemOpen
  \bibfield  {author} {\bibinfo {author} {\bibfnamefont {J.}~\bibnamefont
  {Chelikowsky}}, \bibinfo {author} {\bibfnamefont {D.}~\bibnamefont {Chadi}},
  \ and\ \bibinfo {author} {\bibfnamefont {M.~L.}\ \bibnamefont {Cohen}},\
  }\href@noop {} {\bibfield  {journal} {\bibinfo  {journal} {Phys. Rev. B}\
  }\textbf {\bibinfo {volume} {8}},\ \bibinfo {pages} {2786} (\bibinfo {year}
  {1973})}\BibitemShut {NoStop}%
\bibitem [{\citenamefont {Adachi}\ and\ \citenamefont
  {Taguchi}(1991)}]{Adachi1991}%
  \BibitemOpen
  \bibfield  {author} {\bibinfo {author} {\bibfnamefont {S.}~\bibnamefont
  {Adachi}}\ and\ \bibinfo {author} {\bibfnamefont {T.}~\bibnamefont
  {Taguchi}},\ }\href {\doibase 10.1103/PhysRevB.43.9569} {\bibfield  {journal}
  {\bibinfo  {journal} {Phys. Rev. B}\ }\textbf {\bibinfo {volume} {43}},\
  \bibinfo {pages} {9569} (\bibinfo {year} {1991})}\BibitemShut {NoStop}%
\end{thebibliography}%

\end{document}